# $G^2$ Transition curve using Quartic Bezier Curve

Azhar Ahmad[α], R.Gobithasan[γ], Jamaluddin Md.Ali[β],
[α]Dept. of Mathematics, Sultan Idris University of Education, 35900 Tanjung Malim, Perak, M'sia.
[γ]Dept of Mathematics, Universiti Malaysia Terengganu, 21030, Kuala Terengganu, M'sia.
[β]School of Mathematical Sciences, University Sains Malaysia, 11800 Minden, Penang, M'sia.
{azhar_ahmad@upsi.edu.my, gobithasan@umt.edu.my, jamaluma@cs.usm.my}

**Abstract**
*A method to construct transition curves using a family of the quartic Bezier spiral is described. The transition curves discussed are S-shape and C-shape of $G^2$ contact, between two separated circles. A spiral is a curve of monotone increasing or monotone decreasing curvature of one sign. Thus, a spiral cannot have an inflection point or curvature extreme. The family of quartic Bezier spiral form which is introduced has more degrees of freedom and will give a better approximation. It is proved that the methods of constructing transition curves can be simplified by the transformation process and the ratio of two radii has no restriction, which extends the application area, and it gives a family of transition curves that allow more flexible curve designs.*

## 1. Introduction

In various fields of Computer Aided Geometric Design (CAGD) one of the interests lies in constructing curves and surfaces that satisfy aesthetic requirements. Fairness, or smoothness is an important entity of curve and surface, it is often termed as geometric continuity, $G^k$ or parametric continuity, $C^k$. A generally accepted mathematical criterion for a curve to be fair is that it should have as few curvature extrema as possible. It is desirable that the curvature extreme occur only where the designer wants them [1].

$G^2$ Transition curves of two separated circles, composed of the single segment or a pair of spiral segments are useful for several Computer Graphics and CAD applications. The practical application are e.g., in highway design, railway route, satellite path, robot trajectories, or aesthetic applications [2]. The importance of this design feature is discussed in [3,4]. Spirals have several advantages of containing neither inflection points, singularities and nor curvature extrema. Such curves are suitable for the transition curve between two circles. One of the significant approaches to achieve the transition curve of monotone curvature of constant sign is by using parametric polynomial representation.

One of the most important mathematical representations of curves and surfaces used in computer graphics and CAD is Bezier curves. Their popularity is due to the fact that, they possess a number of mathematical properties which enable their manipulation and analysis [5]. Cubic curves form provide a greater range of shapes that allow the curve to have cusps, loops, and up to two inflection points. This flexibility makes it suitable for the composition of $G^2$ blending curves. Unfortunately their fairness is not guaranteed [8]. Walton and Meek [6,7] have considered a pair of cubic Bezier spiral segment and a planar Pythagorean hodograph quintic spiral forms to blend transition curves of joining circular arcs and straight line segment, to produces fair curves. Five cases of $G^2$ transition curve have been discussed. More improvement is shown in [10], by increasing the degree of freedom of cubic Bezier spiral. Habib and Sakai [2] have also considered a cubic Bezier spiral and suggested a scheme to better smoothness and more degree of freedoms.

This paper introduces a planar quartic Bezier spiral and proposes a method to construct $G^2$ transition curves between two separated circles by composing a pair of spiral segment. We have discussed S-shape and C-shape transition. Using quartic instead of cubic means more degrees of freedom. We exploit the extra degrees of freedom to gain the family of transition curves. Although it involves a long and abstruse mathematical manipulation, the use of symbolic manipulator will be of a great help. This quartic Bezier spiral forms contain neither inflection points, singularities nor curvature extrema. Transition curves can be generated for any two given circles, in other words, it has no limitation on the ratio of two given radii. Also, we introduced a method to find the point of connecting the two spirals that gives a fair $G^2$ transition. A $G^2$ point of contact of two curves is a point where the two curves meet and where their unit tangent vectors as well as their curvatures are matched. We will derive the necessary condition and constrains for composing a spiral and also provide some numerical examples to support the theoretical results.

The remaining part of this paper is organized as follows. Section 2 gives a brief discussion of background, notation and convention. We derive the quartic Bezier spiral in Section 3. In Section 4, a result of transition curves between a point and a circle is presented. Our main result is shown in Section 5. Some numerical examples are showed in Section 6.

## 2. Preliminaries

The following notation and conventions are used. We consider the dot product of two vectors, A and B is given as $A \bullet B$. The notation of $A \times B$ represents the outer product of two plane vectors $A$ and $B$. Note: the dot and outer product results are $A \bullet B = \|A\|\|B\|Cos\theta$ and $A \times B = \|A\|\|B\|Sin\theta$, where $\theta$ is the turning angle from $A$ to $B$. Positive angles are measured anti-clockwise. The norm or length of a vector $A$ is $\|A\|$. In this paper we denoted $R(t)$ as quartic curve. If $T$ is the unit tangent vector to $R(t)$ at $t$, it is denoted as $T = R(t)/\|R(t)\|$ and $\|T\| = 1$. The unit normal vector $N$ to $R(t)$ at $t$ is perpendicular to $T$ and the angle measured anti-clockwise from $T$ is $\pi/2$. The signed curvature of a plane curve $R(t)$ is

$$\kappa(t) = \frac{R'(t) \times R''(t)}{\|R'(t)\|^3} \qquad (1)$$

The signed radius is the reciprocal of (1). It is known that $\kappa(t)$ is a positive sign when the curve segment bends to left and it is negative sign if it bends to right at $t$. If $R'(t)$ and $R''(t)$ are first and second derivation of $R(t)$, differencing of (1) yields

$$\kappa'(t) = \frac{v(t)}{\|R'(t)\|^5}. \qquad (2)$$

where

$$v(t) = \{R'(t) \bullet R'(t)\}\frac{d}{dt}\{R'(t) \times R''(t)\} \\ - 3\{R'(t) \times R''(t)\}\{R'(t) \bullet R''(t)\} \qquad (3)$$

## 3. A planar quartic Bezier spiral

First, we consider a standard quartic Bezier curve

$$R(t) = \sum_{i=0}^{4} B_i C_i \qquad 0 \le t \le 1 \qquad (4)$$

where $B_i$, $i = 0,1,2,3,4$ are control points. And basis functions $C_i$ in parameter $t$ are Benstein polynomial given as

$$C_i = \binom{4}{i}(1-t)^{4-i} t^i, \qquad i = 0,1,2,3,4 \qquad (5)$$

*Theorem 1*

Given a beginning point, $B_0$, and two unit tangent vectors $T_0$ and $T_1$ at beginning and ending points, respectively. Denoted $\theta$ as the anti-clockwise angle from $T_0$ to $T_1$. And ending curvature value given as $\frac{1}{r}$, it is assumed that the centre of the circle of curvature at ending point is to the left of the direction of $T_1$ and in positive value, i.e., $r > 0$. If the control points are given as

$$B_1 = B_0 - \frac{r\rho_0(2\rho_1 + 3\alpha_0(4\rho_1 - 3))^2 \, Sec\theta \, Tan\theta}{108\alpha_0^3 (\rho_0 - 1)\rho_1^2} T_0$$

$$B_2 = B_1 - \frac{r(-1+\alpha_0)(2\rho_1 + 3\alpha_1(4\rho_1-3))^2 \, Sec\theta \, Tan\theta}{108\alpha_0^3 \rho_1^2} T_0 \qquad (6)$$

$$B_3 = B_2 - \frac{r(2\rho_1+3\alpha_0(4\rho_1-3))\begin{pmatrix}-9\alpha_0 Cos\theta(-1+\rho_1)T_1 \\ +(2\rho_1+3\alpha_0(4\rho_1-3))T_0\end{pmatrix} Sec\theta Tan\theta}{108\alpha_0^2 \rho_1^2}$$

$$B_4 = B_3 - \frac{r(2\rho_1 + 3\alpha_0(4\rho_1 - 3)) \, Tan\theta}{12\alpha_0 \rho_1} T_{-1}$$

Where $\alpha_0, \rho_0, \rho_1$ are positive arbitraries given as:

$$\rho_1 = \frac{9}{14}, \qquad \alpha_0 \le \frac{8}{25}$$

$$\rho_0 = \frac{15(1-\alpha_0-\rho_1+\alpha_0\rho_1)}{27-15\alpha_0-26\rho_1+15\alpha_0\rho_1}. \qquad (7)$$

Then $R(t)$ as given by Eq. (4) and (5) is a spiral.
See Appendix 1 for proof.

This quartic Bezier spiral has the following properties;

$R(0) = B_0$, $R(1) = B_4$, $\frac{R'(0)}{\|R'(0)\|} = T_0$, $\frac{R'(1)}{\|R'(1)\|} = T_1$,

$\kappa(0) = 0$, $\kappa(1) = \frac{1}{r}$, $\kappa'(1) = 0$, $\kappa''(0) = 0$ and $\kappa(t) \ne 0$

for $0 < t \le 1$.

Observe that the standard quartic Bezier has ten degrees of freedom. Whereby the quartic Bezier spiral has seven degrees of freedom: one for each of $\theta, r, \alpha_0$ and two of $T_0, B_0$. The value $\rho_0$ is fixed, $\rho_1$ is depending on $\alpha_0$ and $\rho_0$. And $T_1$ had been controlled by $\theta$ and $T_0$.

## 4. A transition curve between a point and a circle

The following are the construction of a transition curve between a point and a circle whereby the result is useful for subsequent sections. There are two possible solutions for this case; referring to curvature of ending point. Fig.1 shows an example of a transition curve between a point and a circle, where $\kappa(1) = \frac{1}{r}$, $r > 0$. We can arbitrarily select one solution since an analogous end result can be obtained for the other one.

For a beginning point $B_0$ and a circle $\Omega_0$ centred at $C_0$ with radius $r > 0$. Denoted that $G_0 = C_0 - B_0$, $\ell = |G_0|$ and $T_0$ is unit vector at $R(0)$. Turning angle, $\theta$, is the angle from $T_0$ to $T_1$. $N_0$ and $N_1$ are unit normal vectors of $T_0$ and $T_1$, respectively. By convention, the unit normal vector at $R(t)$ are on the left side of the tangent at $R(t)$.

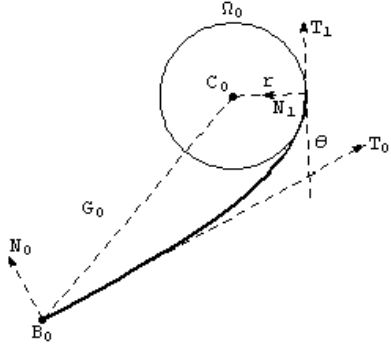

**Figure 1: A transition curve between a point and a circle**

From Theorem 1, we can write $R(1)$ in terms of given unit tangent vectors as
$$R(1) = B_0 + a_0 T_0 + b_0 T_1 \tag{8}$$
Where for $i = 0$,
$$a_i = -\frac{4\rho_1^2 r_i H^2 Sec\theta Tan\theta}{3\alpha_0 (-1 + \rho_0)}, \quad b_i = r_i H\, Tan\theta, \tag{9}$$
$$H = \frac{(2\rho_1 + 3\alpha_0 (-3 + 4\rho_1))}{12\alpha_0 \rho_1^2}. \tag{10}$$
From Fig. 1, we can write
$$G_0 - rN_1 = a_0 T_0 + b_0 T_1 \tag{11}$$
Eq. (11) in the terms of $T_0$ and $N_0$, after simplification using $T_0 \bullet T_1 = Cos\theta$, $T_0 \bullet N_1 = -Sin\theta$, $N_0 \bullet T_1 = Sin\theta$, $N_0 \bullet N_1 = Cos\theta$, (11) can be written as:
$$(G_0 \bullet T_0) T_0 + (G_0 \bullet N_0) N_0 - (-rSin\theta T_0 + rCos\theta N_0)$$
$$= a_0 T_0 + (b_0 Cos\theta T_0 + b_0 Sin\theta N_0) \tag{12}$$
By comparing the coefficients of $T_0$ and $N_0$ from (12), followed by squaring and summing of results, we obtained
$$\|G_0\|^2 = a_0^2 + b_0^2 + r^2 + 2a_0 (b_0 Cos\theta - rSin\theta) \tag{13}$$
Substitution of (9-10) into (13) and by using $\rho_0, \rho_1$ from Theorem 1, such that the points of contact are $G^2$. And $\ell$, $r$, $\theta$, $\alpha_0$ satisfies

$$(q(\theta, \alpha_0)) \ell^2 -$$
$$r^2 \left(1 + \frac{\left(\begin{array}{c}(-1+\alpha_0)^2 \\ \left(\begin{array}{c}2304 - 7008\alpha_0 + 23185\alpha_0^2 \\ -86772\alpha_0^3 + 65646\alpha_0^4\end{array}\right)\end{array}\right)}{385641\alpha_0^6} \Upsilon + \frac{(-1+\alpha_0)^4 (48 - 25\alpha_0)^2}{385641\alpha_0^6} \Upsilon^2\right) = 0 \tag{14}$$

where $\Upsilon = Tan^2\theta$. We can show the condition of this segment is $G^2$ contacts as follows.
$q(\theta, \alpha_0) \to \ell^2 - r^2 > 0$ for $\theta \to 0$ that if $\ell \geq r$, and
$q(\theta, \alpha_0) \to -\infty$ $(< 0)$ for $\theta \to \frac{\pi}{2}$.

The following theorem provides the necessary condition to gain the spiral segment between a point and a circle.

*Theorem 2*
Given a beginning point $B_0$ and a circle $\Omega_0$ centred at $C_0$ with radius $r > 0$. Let $G_0 = C_0 - B_0$ and $T_0$ is unit vector at beginning point. Denoted $\ell = \|G_0\|$ and $\theta$ is the turning angle from $T_0$ to $T_1$. If $\ell \geq r$, then a point and a circle can be joined by a quartic Bezier spiral with $G^2$ contacts on circle.

## 5. A transition curve between two separated circles

In these cases, we used a pair of the quartic Bezier spiral as stated in Theorem 1, denoted by $R_0(t)$ and $R_1(t)$. With assumption that these segments make a contact on circles at $\Omega_0, \Omega_1$ at $R_0(1), R_1(1)$, respectively. $C_0$ and $C_1$ are centre of the circles, with radii $r_0$ and $r_1$. Our interest is to construct a fair transition curve that satisfies following criterion; First, both of the spirals are connected at $t = 0$ with zero curvature, i.e., $\kappa_0(0) = \kappa_1(0) = 0$. Second, both the spirals $R_i(t)$, $i = 0,1$ have the same turning angle, $\theta$. Our focus is to have two spirals that are alike, even though it is different in size and position. Therefore, precise beginning point $B_0$ is required. It is shown that the coordinate of $B_0$ depends on the ratio of the two separated circle and the distance between them.

Referring to S-shape, $B_0$ is taken as a point interpolated by line segment $C_0 C_1$ and its position is determined by the ratio of the two circles, $B_0 = \frac{r_0 C_1 + r_1 C_0}{r_0 + r_1}$. In the case of C-shape, $B_0$ is taken as a point that satisfies $\frac{\|C_0 - B_0\|}{\|C_1 - B_0\|} = \frac{|r_0|}{|r_1|}$, and $C_0 B_0 C_1$ is not collinear. To draw the complete transition curve, two methods have been suggested. First, generate both $R_0(t)$ and $R_1(t)$ separately. Second, generate either $R_0(t)$ or $R_1(t)$ and then by doing the transformation process i.e., translation, rigid rotation or reflection, and uniform scaling over one spiral to gain the other spiral. Such transformation will not change the numbers of curvature extremum and their relative position [10].

Observe that there are two possible solutions for C-shape and S-shape. The analysis will focus on the curve in each case where initial curvature is positive since the opposite case can be defined analogously. Our discussions are based upon Fig.2 and Fig.3.

## 5.1. S-shaped transition curve

Assuming this two segments meet together at $R_0(0) = R_1(0)$ when $t = 0$. Our analysis starts with the finding of equations of $R_0(t), R_1(t)$ when they meet $\Omega_0$, $\Omega_1$, respectively, at $t = 1$. Observe that both curvature at the contact points are positive so $r_0, r_1 > 0$. From Fig.1, this is defined as:

$$G_0 - r N_1 = a_0 T_0 + b_0 T_1 \tag{15}$$
$$G_1 - r_1 M_1 = a_1 F_0 + b_1 F_1 \tag{16}$$

where $N_1$, $M_1$ are unit normal vector of $T_1$ and $F_1$. And $a_i, b_i$ for $i = 0, 1$ are as given in (9).

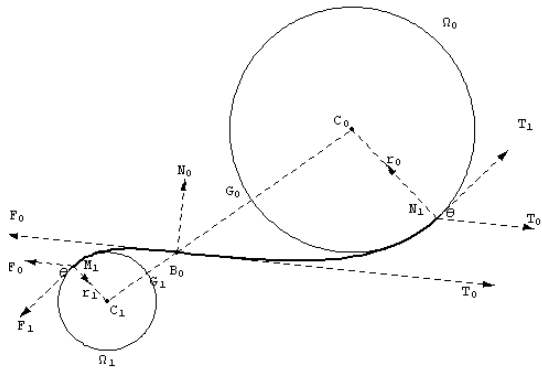

**Figure 2: S-shape transition curve**

According to first criteria, if $R_0(0) = R_1(0)$ and $\kappa_0(0) = \kappa_1(0) = 0$ then $F_0 = -T_0$. For second criteria, if the two spirals had the same turning angle, $\theta$, then $F_1 = -T_1$ and $M_1 = -N_1$. Subtracting (1) from (3) gives us

$$G_1 - G_0 = -(a_1 + a_0) T_0 - (b_1 + b_0) T_1 - (r_1 + r_0) N_1 \tag{17}$$

The result of squaring and summing the components in the terms of $T_0$ and $N_0$ from (6) is

$$\|G_1 - G_0\|^2 = (r_1 + r_0)^2 (f_1^2 + f_2^2) \tag{18}$$

Where

$$f_1 = \cos\theta + H \sin\theta \tan\theta \tag{19}$$

$$f_2 = \frac{\begin{pmatrix} 3\alpha_0 (-1 + \rho_0)(-1 + H) \sin\theta \\ -4\rho_1^2 H^2 \sec\theta \tan\theta \end{pmatrix}}{3\alpha_0 (-1 + \rho_0)} \tag{20}$$

The pair of quartic is obtained upon solving of (18). Observe that the degrees of freedom of quartic Bezier spiral have reduced to two, referring to $\alpha_0$ and $\theta$. Next is to determine $B_0$, followed by $T_0, T_1$ from components of (17). Each value of $\theta$ has unique values of $\alpha_0$. We illustrate it by: $q(\theta, \alpha_0) = (r_1 + r_0)^2 (f_1^2 + f_2^2) - n^2$ where $n^2 = \|G_1 - G_0\|^2 = \|C_1 - C_0\|^2$, so if $0 \le \alpha_0 \le \frac{8}{25}$
$q(\theta, \alpha_0) \to (r_1 + r_0)^2 - n^2 < 0$ if $|r_1 + r_0| < n$ for $\theta \to 0$ and $q(\theta, \alpha_0) \to \infty (> 0)$ for $\theta \to \pi/2$.

*Theorem 3*
Given two circles $\Omega_0, \Omega_1$ centred $C_0$, $C_1$ with radii $r_0, r_1 > 0$. Let $G_i = C_i - B_0, i = 0, 1$. If $|r_1 + r_0| < \|C_1 - C_0\|$, then the two circles can be joined by a pair of quartic Bezier spiral forming S-shaped curve such that all points of contact are $G^2$.

## 5.2. C-shaped transition curve

Analogous to the previous case, the two quartic Bezier spirals for C-shape form are obtained by the following scheme. From Fig.3, these segments meet together at $R_0(0) = R_1(0)$ and make a contact on circles $\Omega_0, \Omega_1$ at $R_0(1), R_1(1)$, respectively when $t = 1$. Observe that the curvatures of contact points are in different sign, $r_{01} > 0$ and $r_1 < 0$. Define $R_0(t)$ and $R_1(t)$ at $t = 1$ as (8-10). According to the corresponding criteria, (16) can therefore be written as

$$G_1 - r_1 M_1 = -a_1 T_0 - b_1 F_1 \tag{21}$$

Where

$$T_0 \bullet T_1 = -T_0 \bullet F_1 = \cos\theta$$
$$T_0 \bullet N_1 = -T_0 \bullet M_1 = -\sin\theta \tag{22}$$
$$N_0 \bullet T_1 = N_0 \bullet F_1 = \sin\theta$$
$$N_0 \bullet N_1 = N_0 \bullet M_1 = \cos\theta$$

Squaring and summing the result in terms of coefficient of vectors $T_0$ and $N_0$ from the difference between with (20) and (15), yields

$$\|G_1 - G_0\|^2 = f_1^2 (r_1 - r_0)^2 + f_2^2 (r_1 + r_0)^2 \tag{23}$$

Where $f_1, f_2$ are similar as (19) and (20), and $\|G_1 - G_0\|^2 = \|C_1 - C_0\|^2$.

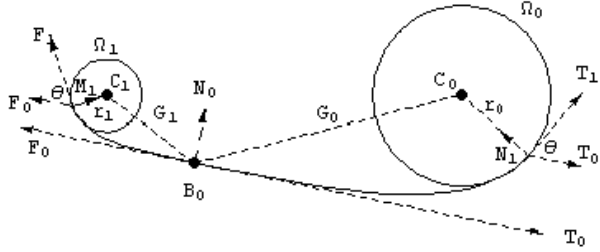

**Figure 3: C-shape transition curve**

The solution for the two quartic Bezier spiral that form C-shaped is obtained by solving (22) for $\alpha_0$ and $\theta$. This is followed by determining the beginning point $B_0$ from (14), $T_0, T_1$ from the difference between (20) and (15). Hence, the following theorem defines the necessary condition for forming C-shape transition curve. Let $q(\theta, \alpha_0) = f_1^2 (r_1 - r_0)^2 + f_2^2 (r_1 + r_0)^2 - n^2$ where $n^2 = \|G_1 - G_0\|^2 = \|C_1 - C_0\|^2$, so if $0 \leq \alpha_0 \leq \frac{8}{25}$

$q(\theta, \alpha_0) \to (r_1 - r_0)^2 - n^2 < 0$ if $|r_1 - r_0| < n$ for $\theta \to 0$ and $q(\theta, \alpha_0) \to \infty (>0)$ for $\theta \to \pi/2$. Hence, we gain the following theorem

*Theorem 4*
Given two circles $\Omega_0, \Omega_1$ centred at $C_0$, $C_1$ with radii $r_0 > 0, r_1 < 0$. Let $G_i = C_i - B_0$, $i = 0,1$. If $|r_1 - r_0| < \|C_1 - C_0\|$, then the two circles can be joined by a pair of quartic Bezier spiral forming C-shaped curve in such a way that all points of contact are $G^2$.

## 6. Numerical Examples

The first example, shown in Fig. 1, represents a transition curve between a point at $(0,0)$ and a circle that is centred at $(13, \sqrt{231})$ with the radius of 5 units. With $\alpha_0 = 0.32$ we obtained $\theta = 1.11088$, followed by $T_0 = (0.880061, 0.474861)$, $T_1 = (-0.0348843, 0.999391)$.

The second example, shown in Fig. 2, represents a S-shaped transition curve between two circles that centred at $(10,7),(0,0)$ with radius 5 and 2 units, respectively. With $\alpha_0 = 0.32$, $\theta = 0.867967$, followed by $B_0 = (2.857142, 2)$, $T_0 = (0.994621, -0.103585)$, $T_1 = (0.721939, 0.691957)$.

The third example, shown in Fig. 3, represents a C-shaped transition curves between two circles that centred at $(20,0),(0,0)$ with radius 5 and 2 units, respectively. With $\alpha_0 = 0.32$, $\theta = 0.867967$ we obtained

$B_0 = (4.956289, -3.723369)$, $T_0 = (0.979901, -0.199486)$, $T_1 = (0.662711, 0.748876)$ and $F_1 = (-0.317190, 0.948362)$

## Conclusions

It has been demonstrated that fair curves can be designed interactively using quartic Bezier spirals. Since this Bezier quartic also has NURBS representations, curves designed using a combination of quadratic, cubic, quartic spirals, circular arcs and the straight line segment can be represented entirely by NURBS. This quartic Bezier spiral is more flexible than cubic Bezier spiral in general because it has seven degrees of freedom; this will give a family of transition curves in either S-shape or C-shape forms and the spiral can be generated from two arbitrary circles. The advantage of using this is the ratio of two radii of separated circles has no restriction. By using the precise beginning point for those spiral segments as suggested. In this paper, we can also simplify the computation through transformation process.

# Appendix

*Proof of Theorem 1*

This proof is based on constructive prove. Which it is start by finding $R(t), R'(t), R''(t)$ and $R'''(t)$, followed by $R'(t) \times R''(t), R'(t) \bullet R'(t), R'(t) \times R'''(t)$ and $R'(t) \bullet R''(t)$. Substitute the results into (3) followed by into (2). Hence, after replacing $\rho_0, \rho_1$ from (4), the result is

$$\kappa'(t) = \frac{85698\alpha_0^5 \sum_{i=0}^{9} P_i(1-t)^{9-i} t^i}{\sqrt{\left(r^2 \left(M \, Tan^2\theta + N\right)^3\right)}} \qquad (24)$$

where

$$M = 4(-1+t)^2(-1+\alpha_0) \qquad (25)$$
$$\left(25+19t-44t^2+\left(-25-19t+113t^2\right)\alpha_0\right)^2$$

$$N = \begin{pmatrix} 2(-1+t)^2(25+44t) \\ +\left(-100+24t+390t^2-314t^3\right)\alpha_0 \\ +\left(50-12t+81t^2+88t^3\right)\alpha_0^2 \end{pmatrix}^2 \qquad (26)$$

$P_i$ are as shown below

$$P_0 = -312500 Sec^2\theta(-1+\alpha_0)^5,$$
$$P_1 = -2437500 Sec^2\theta(-1+\alpha_0)^5,$$
$$P_2 = -3000(-1+\alpha_0)^3 H_2,$$
$$P_3 = -4140(-1+\alpha_0)^3 H_3, \qquad (27)$$
$$P_4 = -1035(-1+\alpha_0) H_4,$$
$$P_5 = -3105(-1+\alpha_0) H_5,$$
$$P_6 = -28566\alpha_0(-1+\alpha_0) H_6, \, P_7 = 42849\alpha_0^2 H_7,$$
$$P_8 = 12854\alpha_0^3 H_8, \, P_9 = 0.$$

with

$$H_2 = 3836 - 8822\alpha_0 + 2111\alpha_0^2$$
$$+ 2(-1+\alpha_0)(-1918+2493\alpha_0) Tan^2\theta$$
$$H_3 = 7674 - 18823\alpha_0 - 351\alpha_0^2$$
$$+ (-1+\alpha_0)(-7674+11149\alpha_0) Tan^2\theta$$
$$H_4 = \begin{pmatrix} 38916 - 154084\alpha_0 + 161956\alpha_0^2 \\ -74824\alpha_0^3 - 43839\alpha_0^4 \end{pmatrix} +$$
$$4(-1+\alpha_0)^2(9729-19063\alpha_0+6459\alpha_0^2) Tan^2\theta$$
$$H_5 = 6348 - 11776\alpha_0 - 3760\alpha_0^2$$
$$-31304\alpha_0^3 - 40583\alpha_0^4 \qquad (28)$$
$$+ 92(-1+\alpha_0)^2(69+10\alpha_0-150\alpha_0^2) Tan^2\theta$$
$$H_6 = 5(276-373\alpha_0-900\alpha_0^2-1338\alpha_0^3)$$
$$+ (-1+\alpha_0)^2(1380-1373\alpha_0) Tan^2\theta$$
$$H_7 = 5(230-614\alpha_0-514\alpha_0^2+783\alpha_0^3)$$
$$+ 46(-1+\alpha_0)^2(13-15\alpha_0) Tan^2\theta$$
$$H_8 = 5(46-176\alpha_0+107\alpha_0^2) + 46(-1+\alpha_0)^2 Tan^2\theta.$$

For $P_i = \varphi(\alpha_0) H_i$, $i=2,3,...,8$, since $0 < \theta < \pi/2$, $0 < \alpha_0 < 1$, therefore $\varphi(\alpha_0)$ are positive. So $H_i$ should be positive in values for $P_i > 0$. For each $H_i > 0$, every polynomial, $\psi(\alpha_0)$ in terms of $\alpha_0$ should satisfy the following inequality

$$H_2: 3836 - 8822\alpha_0 + 2111\alpha_0^2 > 0,$$
$$(-1918+2493\alpha_0) < 0$$
$$H_3: 7674 - 18823\alpha_0 - 351\alpha_0^2 > 0,$$
$$(-7674+11149\alpha_0) < 0$$
$$H_4: \begin{pmatrix} 38916-154084\alpha_0+161956\alpha_0^2 \\ -74824\alpha_0^3-43839\alpha_0^4 \end{pmatrix} > 0,$$
$$(9729-19063\alpha_0+6459\alpha_0^2) > 0$$
$$H_5: \begin{pmatrix} 6348-11776\alpha_0-3760\alpha_0^2 \\ -31304\alpha_0^3-40583\alpha_0^4 \end{pmatrix} > 0,$$
$$(69+10\alpha_0-150\alpha_0^2) > 0$$
$$(29)$$
$$H_6: (276-373\alpha_0-900\alpha_0^2-1338\alpha_0^3) > 0,$$
$$(1380-1373\alpha_0) > 0$$
$$H_7: (230-614\alpha_0-514\alpha_0^2+783\alpha_0^3) > 0,$$
$$(13-15\alpha_0) > 0$$
$$H_8: (46-176\alpha_0+107\alpha_0^2) > 0$$

$H_i$ are positive if $0 \le \alpha_0 \le \alpha_{min} \left(= Min[\psi(\alpha_0)]\right)$. As the result, from $H_7$ we obtain $\alpha_{min} = 0.325958 \approx \frac{8}{25}$. This ends of the proof.